# Combining genetic algorithm and compressed sensing for features and operators selection in symbolic regression

A. Mazheika[1], S.V. Levchenko[2], and L.M. Ghiringhelli[3,4],

1) BasCat – UniCat BASF JointLab, Technische Universität Berlin, Germany
2) Skolkovo Institute of Science and Technology, Moscow, Russia
3) Department of Materials Science and Engineering, Friedrich-Alexander-Universität Erlangen-Nürnberg, Germany
4) Physics Department and IRIS-Adlershof, Humboldt Universität zu Berlin, Germany

Abstract

Symbolic-inference methods have recently found a broad application in materials science. In particular, the Sure-Independence Screening and Sparsifying Operator (SISSO) performs symbolic regression and classification by adopting compressed sensing for the selection of an optimized subset of features and mathematical operators out of a given set of candidates. However, SISSO becomes computationally unpractical when the set of candidate features and operators exceeds the size of few tens. In the present work, we combine SISSO with a genetic algorithm (GA) for the global search of the optimal subset of features and operators. We demonstrate that GA-SISSO efficiently finds more accurate predictive models than the original SISSO, due to the possibility to access a larger input feature and operator space. GA-SISSO was applied for the search of the model for the prediction of carbon-dioxide adsorption energies on semiconductor oxides. The obtained with GA-SISSO model has much higher accuracy compared to models previously discussed in the literature (based solely on the O 2$p$-band center). The analysis of features importance shows that, besides the O 2$p$-band center, the contribution of the electrostatic potential above adsorption sites and the surface formation energies are also important.

**Introduction**

Symbolic inference (therein including regression and classification tasks) is a class of statistical-learning methods in which explanatory models for a given data set are learned as algebraic or Boolean expressions containing data features, mathematical or logical operators, functions, and coefficients [1,2]. In contrast to other popular machine-learning methods such as artificial neural networks, linear regression, or tree-based methods, symbolic inference does not have any predefined model class, i.e., the functional form for the learned model is searched together with the fitting parameters [1]. Symbolic inference (SI) found its application in physical sciences and in particular in materials science, since it provides interpretable mathematical models, which can be easily analyzed. In materials science, SI provides the model for a target materials property as function of so-called *primary features*, i.e., materials-related properties. The primary features that are selected to be part of the SI-learned model, can be interpreted as the *material's genes* [3] that trigger, facilitate, or hinder the target property. SI found its application in the prediction of the crystal structure of solid materials [6], lattice parameters and bulk moduli [7], superconductivity [4], topological insulators [5], Gibbs free energies of inorganic materials [8], adsorption properties of hydrogen [10] and other molecules [9].

Recently, some of us developed a method which combines symbolic inference with compressed sensing – the Sure-Independence Screening and Sparsifying Operator [6,11]. SISSO expresses its learned models as linear combinations of *generated* features. Generated features are nonlinear functions of the input (*primary*) features, i.e. they are combinations of input features and mathematical operators (summation, multiplication, powers, exponentials, etc.). The maximum number of unique primary features, which a generated feature can contain, is the *complexity* ($\Omega$) of the model. In the so-called *feature-creation step* of the SISSO algorithm [12], the construction of the generated features is done recursively, so that at each step more complex features are built from a pool of less complex features generated at all previous iterations including primary ones. Next, in the *descriptor-identification* step the SISSO model is created as a linear combination of selected generated features. The learning of a SISSO model is done via minimization of the (square) difference between actual target property ($y$) and its SISSO counterpart ($y^{\text{SISSO}}$). For imposing the sparse solutions a $l_0$-regularization parameter is introduced, which is equal to the number of linear terms, i.e., the dimensionality ($D$) of the model. Since the number of all possible complex features grows combinatorially, it becomes impractical to consider all possible combinations, even for relatively small sets of primary features and SISSO hyperparameters ($D$, $\Omega$). To enable the model identification a sure-independent-screening (SIS) approach is used [13]. The algorithm is iterative, and the number of iterations is equal to the dimensionality $D$. SIS implies that at first iteration only a set of $M$ most correlated complex features to the target property is selected, and on next iterations

– $M$ correlated features to residuals "target property - model obtained on a previous iteration" are selected. After such selection at $D$th iteration least-squares fit is done for all $D$-nomial sets out of $\bigcup_i^D M_i$ complex features, and the one with lowest training error is selected. In the SISSO algorithm a model is selected based on the training accuracy, and the assessment of the prediction ability of the model is done through the cross-validation (CV) procedure. The prediction accuracy depends on hyperparameters. Besides mentioned above dimensionality ($D$) and complexity ($\Omega$), the sets of primary features and of mathematical operators are also the hyperparameters in the SISSO method that need also to be tuned. For given $D$ and $\Omega$, the maximum number of primary features per model is at most $D$ times $\Omega$. If the overall number of all primary features $n$ is larger than the number of features in a SISSO model $D \cdot \Omega$, then the number of all possible combinations of size up to $D\Omega$ grows again combinatorially: $n!/((D\Omega)!(1-D\Omega)!)$. So, far not always features selection can be done in a straightforward way. The other point is that, a too large set of primary features and operators inside the learned model might lead to overfitting; and in contrast, a SISSO model obtained with a poor set of features and operators might be not flexible enough for accurate predictions – the underfitting. We note that, mainly due to memory restrictions, when $\Omega > 4$ and the number of primary features is more than ~25, it becomes hardly possible with nowadays available machines to run SISSO "in one shot", because the number of generated features explodes is unmanageable. To address the above challenges, we have introduced a stochastic algorithm to search for the optimal subset of primary features and mathematical operators, out of the initially given set of both entities.

The search for the optimal parameters $D$ and $\Omega$ is usually done on a grid. For selection of features several schemes have been proposed. Among them variable importance measure of random forests [14], LASSO [15], minimum redundancy – maximum relevance [16], etc. More recently Regler et.al. proposed total cumulative mutual information estimator [17]. This method is non-parametric and it identifies the sets of primary features which are statistically related to the target property. Guo et.al. developed a variable selection method where at each iterative step a set of features is randomly selected and added to the set of features with best prediction performance obtained so far [18]. This method converges quite fast even for relatively big features sets, but the procedure can easily get stuck into a local minimum.

Also recently another way for tackling the problem of a combinational growth of the number of generated features was proposed. Foppa et.al. developed a hierarchical symbolic regression approach – hiSISSO [7]. There SISSO models are obtained in a cascade-like way. At first step a SISSO model with lower complexity is obtained. At the second step this model is used as an input feature in addition to other features to obtain a final SISSO model with higher complexity. This approach addresses the problem of a large number of complex features, but not the primary features selection for generation of these complex. Below we apply this approach, and herewith we define the complexity and dimensionality of the model-input feature obtained at the first step as $D^{hiSISSO}$ and $\Omega^{hiSISSO}$, and for more complex expression at the second step – $D$ and $\Omega$ as usual.

In this work, we present a method for selection of primary features and mathematical operators based on genetic algorithm (GA), and its particular application for SISSO. Genetic algorithm is a natural way of the search for optimal strings of biological genes in nature. It has found broad application in materials science as a tool for global minimum/maximum search for atomic systems [19,20], in machine learning [21], etc. The main idea is that the selection of parent strings with "good" features (genes, atomic environments, etc.) can deliver new strings containing these fragments so that they can even outperform the parents in terms of the target performance. Sequential generation of new strings follow at some point finding of the one with the best possible performance. In our method, dubbed GA-SISSO, the *genes* are primary features and mathematical operators. Each set of features and operators forms a string with certain SISSO cross-validation root-mean-square error (CV-RMSE). GA algorithm explores the space of features and operators searching for a string with lowest CV-RMSE. In our implementation, we generally follow the scheme previously proposed by Bhattacharya et.al. [19,22]. Crucially, in order to avoid getting stuck in a local minimum, we do not preserve the generations. Selection is done for the whole pool of strings dependent on a fitness function which is related to CV-error value. The validation of our method was done for a set of bulk perovskite materials in particular for prediction of the lattice constant [7].

As novel application, we apply GA-SISSO to the prediction of carbon-dioxide adsorption energies on semiconductor oxides. Carbon dioxide is nowadays considered as a gas playing the main role in "green-house" effect. It is released first of all by the chemical industry and during energy production (for example coil burning, etc.). The decrease of its concentration in the air or at least reduction of its emissions is a central societal issue today. One of such ways is its conversion to fuels or other useful organic chemicals. For this reason different chemical reactions of $CO_2$ reduction are intensively studied by many researchers around the world. The adsorption energies of $CO_2$ play in such processes a central role because of Sabatier principle which states that carbon dioxide should not be bonded too strongly in order the surface of a catalyst was not poisoned, and it should not be bonded too weakly because otherwise the $CO_2$ molecules would easily desorb without undergoing chemical transformations. The influence of bonding energies of carbon dioxide on reaction performance was found to play a major role also in reactions of methane oxidation where carbon dioxide is actually an unwanted byproduct [24,25]. In other chemical processes, the role of $CO_2$ is to hamper the aggregation of surface cations into metallic nanoparticles, that promotes the selectivity [26]. Thus, predictions of $CO_2$ binding energies with underlying surfaces are important in the search for new catalytic materials enabling fast screening of materials in a high-throughput way [25]. Using GA-SISSO we find the models that are significantly more accurate than models fitted to previously proposed popular descriptors [27,28,29].

**Results and discussion**

**Tests of internal GA parameters**

The choice of features has significant influence on prediction ability of the SISSO model. In material science for predictions of quantitative or categorical properties the features are often the atomic properties, properties of bulk materials, their surfaces etc. Dependent on a sampling the number of features can range from several ones to hundreds or even thousands. Although the SISSO algorithm does selection of primary and generated features during the SIS and following SO steps, this choice reflects only the training accuracy, and is not responsible for prediction ability of a corresponding model. The choice of mathematical operators is also important since some operators can follow failures in prediction. For example square root or division do not make sense if features or their combinations are negative or equal to zero. Similar to other hyperparameters selected sets of features and operators have to satisfy the bias-variance trade-off – compromise between under- and overfitting. In the Data Science this compromise is often searched through the cross-validation procedure (CV) with respect to hyperparameters. There are several CV schemes. Quite popular is the $k$-fold CV. Here a data set with $N$ data points is split $k$ times into two subsets of the size $N/k$ and $N(k-1)/k$, so that there is no overlap between subsets. Next, the models obtained for bigger size subsets are tested on smaller sets providing the overall CV-RMSE. For small size data sets, often the leave-one-out CV is used, in which $N$-size sampling is split $N$ times, and for each left sample the model obtained for the rest $N$-1 sampling is tested. These methods have been implemented in our algorithm, and we used leave-one-out CV for the data sets of size of tens samples and k-fold CV for hundreds of samples. For data sets with thousands data points one can test prediction ability splitting the data into a bigger training and a smaller validation subsets. However, in this work we did not use this way since our data sets do not exceed the size of five hundreds data points.

GA-SISSO is performed for fixed $D$ and $\Omega$, and includes several steps:

(0) Generation of an initial pool of strings with randomly chosen features and operators and the calculation of corresponding CV-RMSE. The size of this pool is at least equal the overall number of primary features implying that probability of each feature to appear in any string within this pool is close to one. Before the actual calculation of CV-RMSE values we adopt a preselection step: single standard SISSO job is done for a given string, and next CV-RMSE is calculated for a truncated set of features and operators – only for ones which are selected by single SISSO, i.e. appear in obtained model. Also after this single SISSO the parameter $M$ (the number of most correlated complex features to the target property or to the residuals) is reduced to the highest ranking number of any complex feature in the model of single SISSO job. This results in significant reduction of the space of complex features $\bigcup_i D M_i$ at SO step of SISSO during CV runs. This preselection procedure of features and operators and reduction of *M* provide significant reduction of computational time needed for running k jobs in the case of k-fold CV or N jobs for N-size sampling in the case of leave-one-out CV.

(1) Selection of two strings according to their fitness function with certain probability of fitness-function inversion *inv* (see below).

(2) Random selection of a half of features and a half of operators from each string selected on the previous step, followed by the combination of these two subsets into a new string (so called *crossover*).

(3) *Mutation* step: certain percentage of features and operators is removed from a new string; and at the same time, some new features/operators are added. The amounts of removed/added features/operators are defined by corresponding probabilities – two parameters for features $m_r^f$ and $m_a^f$, and two for operators $m_r^o$ and $m_a^o$ (Table 1). At the same time, the maximum number of possible features in a string *L* is fixed in order to prevent potential "explosion" if the overall number of features is pretty big.

(4) We check if the new string is already in the pool of previously considered strings. If not the case, the CV is run for it in the same way as for randomly generated strings at the step 0 including single SISSO job in the beginning. Next, the pool of all considered strings is extended, and the fitness functions are updated.

Table 1. GA and some SISSO internal parameters and their acronyms.

| acronym | description |
|---|---|
| D | SISSO descriptor dimensionality |
| Ω | Complexity: maximum number of distinct features per dimension |
| M | The amount of most correlated generated features to the target property or to a residual selected at the SIS-step in SISSO algorithm. |
| inv | Probability of fitness function inversion |
| $m_r^f$ | Feature removal probability from a string |
| $m_a^f$ | Feature addition probability to a string |
| $m_r^o$ | Operator removal probability from a string |
| $m_a^o$ | Operator addition probability to a string |
| L | Maximum allowed number of features in a string |

For the fitness function ($f_i$) we used in this work a quadratic dependence of sorted samples according to the rank:

$$f_i = \left( (N_{all} - rank_i) / (N_{all} - 1) \right)^2 \quad (1)$$

here $N_{all}$ is the size of the pool with all strings, and $rank_i$ – position of $i$th string in the list of strings sorted in ascending order according to CV-RMSE. In such a way, all strings in the pool have the values of fitness-function in the range [0, 1]. This fitness-function gives more priority to strings with lower values of CV-RMSE. In the literature quite often the fitness function is used with CV-RMSE instead of rank. In our case, however, the presence of the division operator in operators set results sometimes in huge relative values of CV-RMSE for strings which contain samples with features with close to zero values. This results that corresponding fitness function with CV-RMSE acquires very "skewed" shape (Figure S) that is less suited for operating with random numbers, since the majority of strings has relatively close values of fitness function in contrast to discussed. Selection of parent strings is done as follows: a random string is picked from the pool, and if its fitness-function is higher than a random number uniformly selected in the range [0, 1], it is selected for the next steps. Otherwise the procedure is repeated. We also adapted here the inversion of the fitness function as defined in Ref. 22, so that fitness functions for all strings $f_i$ are inverted to $(1 - f_i)$ with a certain probability *inv*. This guarantees selection of "bad" strings that prevents getting stuck in a local minimum. Fitness-function (1) depends on the pool of strings. So, after adding a new string into the pool, the fitness-function is updated for each of them.

For crossover features and operators are randomly chosen. Their number composes 50% of each parent string if the overall numbers are even. In the case if parent strings contain odd numbers of features (*l*), then from a string with higher value of fitness-function $(l+1)/2$ features are chosen, and $(l-1)/2$ otherwise. The same was implemented for operators.

The performance of our algorithm was tested for a set of bulk ternary perovskites $ABO_3$ with lattice constant as the target property [7]. This data set contains 504 materials where *A* elements are chosen among *s*-elements Li-Cs, Be-Ba, *d*-elements Sc, Y, *f*-elements La-Sm. *B* elements: *d*-elements Ti-Zn, Zr-Cd, Ta, W, Pt, *p*-elements Al, Ga, Ge-Pb, Sb, Bi. The *primary features* set includes atomic properties of gas-phase atoms *A* and *B*: radii of valence and of *s*-orbitals, atomic numbers, HOMO and LUMO values, ionization potentials, electron affinities and electronegativities, number of valence electrons in atoms *A* – overall, 17 features. Here we use this set for benchmarking our GA algorithm.

First, to test the internal parameters of GA with reasonable computational expenses we randomly selected a subset of 30 materials, and benchmarked our method for SISSO two-dimensional models with $\Omega = 4$ using 10-fold CV. The baseline in these tests was a model obtained on top of SISSO CV done with all seventeen features and all mathematical operators (CV-RMSE =

0.057 Å). In GA search we restricted the maximal number of candidate features in a single string to eight ($L = 8$), since this is the maximal possible number of unique features for $D = 2$ and $\Omega = 4$. For each set of parameters we performed 2-3 GA runs. The conclusion that the GA is converged was made when the number of GA iterations done after finding a string with lowest CV-RMSE was about twice as large as the number of iterations needed for finding this string. We note herewith that this criterion requires a lot of computational time, and we used it only for benchmarking tasks. Alternatively one can stop GA at the point when desired RMSE is achieved.

By setting inversion and all mutation probabilities in GA to zero, not a single string corresponding to the baseline is found (Figure 1a). This shows the need of setting non-zero values at least for some of these probabilities. In the next steps, we tested probability of the fitness-function inversion, as well as probabilities of mutation of features and operators. These tests were done for each parameter separately with the values 0.1, 0.25, 0.5 and 1.0 keeping the other probabilities equal to zero. As it is shown in Figure 1b, the best results in terms of finding lowest CV-RMSE string were obtained for $inv = 0.1$, $m_r^f = 0.1$-$0.25$, $m_r^o = 0.5$-$1.0$, $m_a^f = 0.25$-$1.0$, $m_a^o = 0.5$. These settings allow to obtain the strings with CV-RMSE close to or even outperforming the baseline. Regarding inversion probability $inv$, in the study of Bhattacharya [22] the same optimal value was found. Larger values result in too frequent selection of "bad" strings, which destroys the idea of genetic algorithm turning it into a random search. The same is true for high probabilities of selected features removal $m_r^f$. At the same time, we observed that the probability of features inclusion into a new string $m_a^f$ is very important. This means that dealing with features that are selected only via crossover does not guarantee finding the strings with lower CV-RMSE. Addition of remaining features (up to a predefined maximal number of features $L$) is necessary for perturbing the system in order to go beyond these limitations. Regarding mathematical operators, both exclusion of selected and inclusion of remaining ones ($m_r^o$, $m_a^o$) are important and provide significant improvement of obtained strings in terms of CV-RMSE. So, update of operators is very important for preventing getting stuck in a local minimum string.

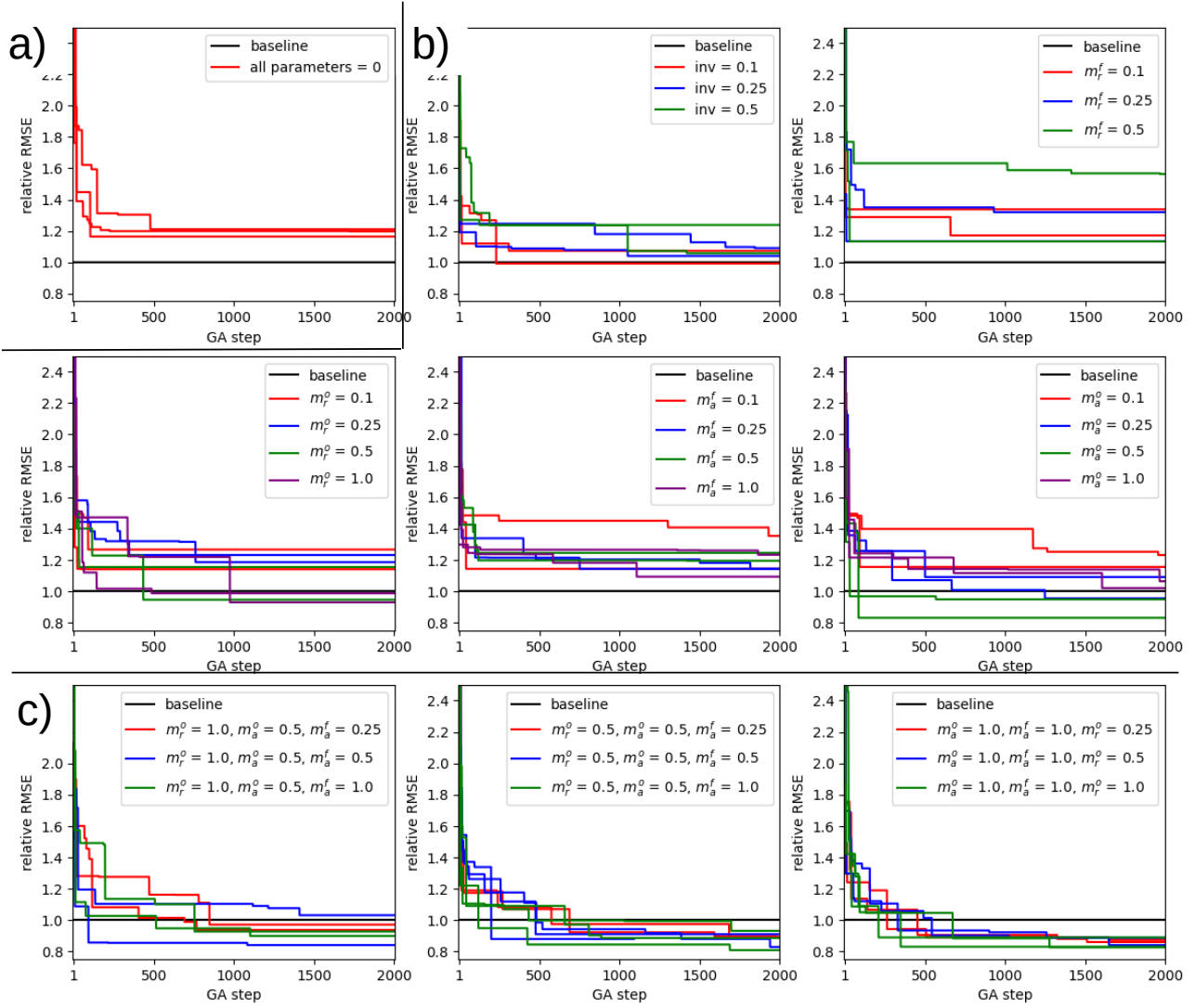

Figure 1. GA performance in terms of CV-RMSE (relative to baseline) dependence on GA steps for SISSO models of lattice constant in perovskite materials. a) Case when all internal GA probabilities are zero. b) Inversion probability or one mutation probability is changed whereas all others are fixed to zero. c) Combinations of different values of $m_a^f$, $m_a^o$ and $m_r^o$ with fixed $m_r^f = 0.1$ and $inv = 0.1$.

On the basis of these results, we next tested the combinations of mutation probabilities. Here, we fixed the probability of removal of selected features to 0.1, and the same value for inversion of fitness-function was fixed. The other probabilities were changed according to the observation described above. Besides outperforming the baseline, we have taken into account in how many iterations the minimum can be found. As one can see in Figure 1c, fixing $m_r^o$ at 1.0 and $m_a^o$ at 0.5, while changing only the probability of features addition $m_a^f$, does not always result in finding better models than the baseline one. At the same time, reduction of $m_r^o$ to 0.5 improves the GA performance. Fixing $m_a^o$ and $m_a^f$ at 1.0 yields the best performance with almost no dependence on probability of the selected operators removal. We also note that the same tests have been carried out for the same dataset but for different dimensionality and complexity of descriptors: $D = 1$ and $\Omega = 6$ (see the SI). We observed very similar trends that indicates that the "rule of thumb" in our GA is to set probabilities for selected features removal $m_r^f$ ~0.1, for unselected features addition $m_a^f$ 1.0 (with certain limit of the maximum number of features in a string $L$), for removal of selected operators $m_r^o$ in the range 0.25 to 0.5, and for addition of remaining operators $m_a^o = 1.0$.

The described above tests were done for a relatively small set of primary features (17). To check the performance of GA algorithm for larger features sets, we have chosen a larger dataset with 100 randomly selected perovskite samples, and extended this dataset with "synthetic" features consisting of random numbers with a uniform distribution. The range of these random numbers was from zero to the rank of a feature in sorted list of these features starting from one. We considered sets of 24, 30, 40, 50, 70 and 100 features. Corresponding tests were performed 2-3 times for each set, all for SISSO models with $D = 2$ and $\Omega = 4$. In Figure 2, the number of GA steps needed to find the SISSO model with lowest CV-RMSE for the different numbers of features is shown. We emphasize here that for each set of features GA-SISSO found always the same model with lowest CV-RMSE despite the "noise" from synthetic features. We clearly see the growth of the number of steps needed to find the searched model with respect to the number of features. The reason is that since the maximal number of considered features per a trial string $L$ is limited (in this case $L = 8$), larger numbers of strings need to be considered for larger sizes of features sets. However, we can conclude that for the data sets with up to seventy features about two thousands GA steps are enough for finding the model with the lowest CV-RMSE.

Finally we have performed the GA search for SISSO models of a lattice parameter in $ABO_3$ perovskite materials for all 504 materials for different complexities as in the study of Foppa et.al. [7]. We considered GA-SISSO models with $D = 5$ and $\Omega = 2, 4$. And in addition, we applied the hiSISSO approach with $D = D^{hiSISSO} = 5$, $\Omega = 4$ and $\Omega^{hiSISSO} = 2$, but in contrast to original work both models at first and second steps were obtained with GA-SISSO. Basically the aim was to compare the results of GA to what is obtained using all 17 features for a single SISSO job and hiSISSO as it was done in the original study. Here, we used the 5-fold cross-validation procedure. The distribution of absolute CV test errors for the best models obtained via GA and for the corresponding single SISSO jobs are shown in Figure 3. The models obtained with GA in all studied cases outperform the ones obtained from single SISSO runs. In the cases of $\Omega = 2$ and 4 the improvement in terms of CV-RMSE was 9 and 6% respectively, in terms of CV mean absolute error (CV-MAE) – 8 and 7%, and CV median absolute error (CV-MedAE) – 9 and 12%. In the case of hiSISSO, where the GA-SISSO model of higher complexity contains a GA-SISSO model of lower complexity, the improvement is pronounced even more significantly – CV-RMSE for 21%, CV-MAE – 22%, and CV-MedAE – 25%.. The corresponding models are shown in the SI. By the way, the CV-RMSE improvement for about 20% was also observed for described above 30 samples data set shown in Figure 1c.

We conclude that GA provides SISSO models with consistently improved prediction accuracy compared to simple SISSO.

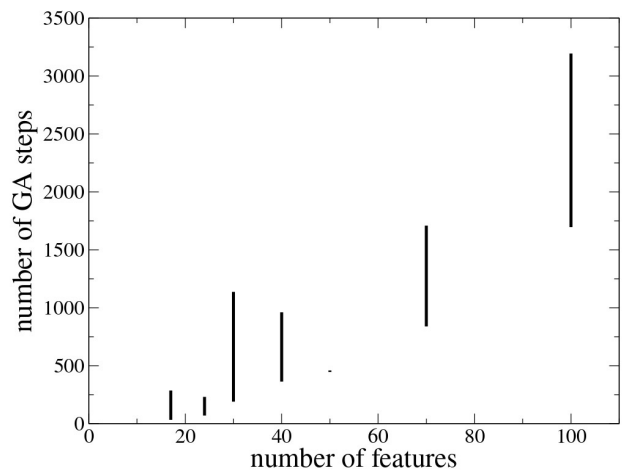

Figure 2. The dependence of the number of GA steps needed to find a string with lowest CV-RMSE on the number of primary features for a set. All sets of primary features include 17 common features and the rest are randomly generated numbers. For each set of features the distribution obtained from several GA runs is shown.

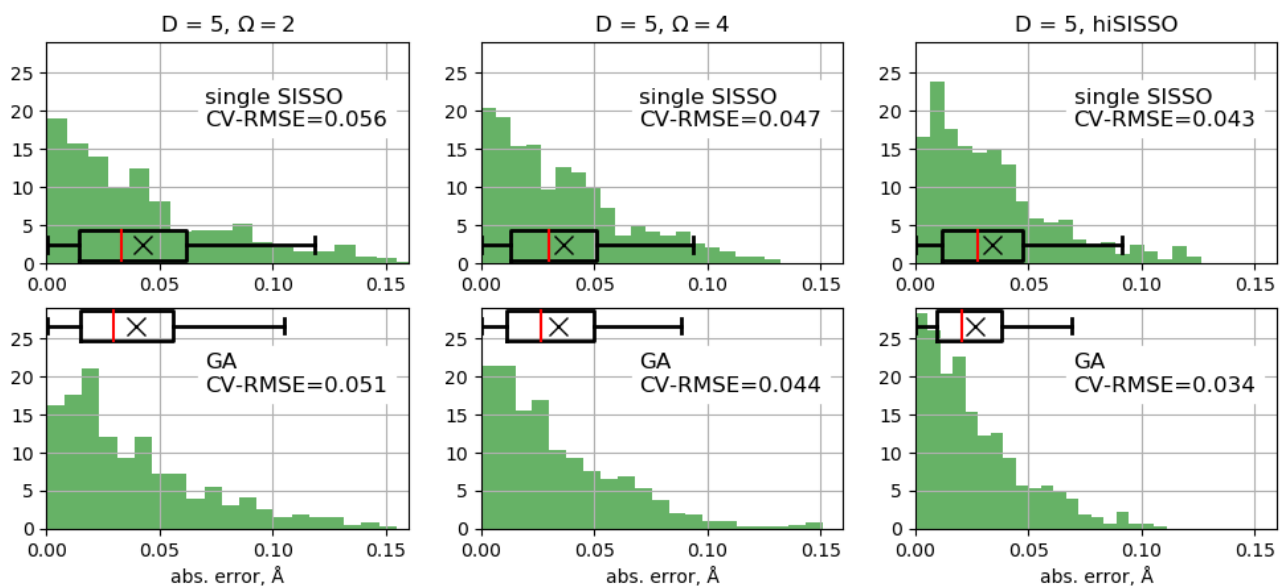

Figure 3. The distribution of absolute test errors for SISSO 5D-models of perovskite lattice vector [7] obtained during 5-fold CV. All models in upper row are obtained applying single SISSO run, in bottom row – using GA-SISSO. In the left column models with complexity $\Omega = 2$ are shown, in middle – $\Omega = 4$, in the right – hiSISSO models with $\Omega = 4$ and an input-feature GA-SISSO model with $D^{hiSISSO} = 5$ and $\Omega^{hiSISSO} = 2$. Green histograms – densities of samples, black boxes – 0, 25, 50,

75 and 95 %, red – median.

**Applications of GA-SISSO for predicting $CO_2$ adsorption energies**

We now present the results of GA-SISSO search for predictive models of carbon dioxide adsorption energies. In several studies, the center of O 2$p$-band ($PC$) was considered as a descriptor for adsorption energies on oxide surfaces [27,28,29,30], similarly to $d$-band center for transition metals and alloys. However, analyzing the data set we discuss below, we found that the lowest 10-fold CV-RMSE in the polynomial model containing only $PC$ was observed for the power degree 4 – 0.47 eV (Figure S). Evidently, prediction accuracy for the model containing only $PC$ is not very high. The reason is that the energy of O 2$p$-band levels is not the only actuating mechanism responsible for carbon dioxide activation. As we have shown in our previous study [23], quite often $CO_2$ molecules are additionally bonded to surface cations – with one or several in the neighborhood of oxygen atoms. This results in low prediction accuracy of the adsorption energy when a model contains only the O 2$p$-band center, and suggests that the properties of surface cations may also have to be taken into account for obtaining accurate prediction models.

To obtain the GA-SISSO models of $CO_2$ adsorption energies on semiconductor oxides, we have used the same training set as in our recent study [23]. It contains 255 samples with $CO_2$ molecules on binary and ternary oxides of Li-Cs, Mg-Ba, Al-In, Si-Sn, Sb, Sc-V, Y-Nb, La and Zn. The range of adsorption energies is quite wide: from very weakly bound molecules with -0.1 eV, to strongly bound $CO_2$ (-3.6 eV). The set of primary features included properties of gas-phase atoms (ionization potentials, electron affinities, etc.), surfaces (formation energies, work functions, band gap, conduction band minimum and surface O-atoms above which $CO_2$ molecules are activated and surface cations (geometric properties, projected density of states moments, etc.) – overall 46 features [23].

To define optimal dimensionality and complexity {$D$, $\Omega$} we performed a grid search in the range from 4 to 7 for dimensionalities and 4 to 8 for complexities. Two subsets of primary features have been selected for these jobs. The first one contained sixteen randomly selected features, and the second one composed of sixteen most correlated features to $CO_2$ adsorption energy. For all these sets we performed 10-fold cross-validation jobs with all mathematical operators (Table 2). Because of the reasons discussed above, the dependence of CV-RMSE on $D$ and $\Omega$ is not smooth. Moreover, for several grid-points we observed the CV-RMSE divergence (mathematically undefined values), and in quite many cases the errors exceed the one observed for polynomial fit of PC indicating the overfitting. Thus, next we performed the search for best SISSO models for those sets of $D$ and $\Omega$ which were found to have lowest CV-RMSE's dealing with any of considered features sets, and those which can finally deliver low CV errors according to expected trends disregarding grid-points where the error divergence had been observed. Relatively low CV-errors were obtained for lower complexities (4 and 5), so we considered the following {D, Ω} sets for genetic algorithm search: {4,4}, {4,5}, {5,4} and {5,5}.

Table 2. 10-fold CV-RMSE (in eV) obtained for single SISSO runs for different dimensionalities (D) and complexities (Ω), for two subsets of primary features: randomly selected and correlated ones to $CO_2$ adsorption energy.

| D | 4 | | | | | 5 | | | | | 6 | | | | | 7 | |
|---|---|---|---|---|---|---|---|---|---|---|---|---|---|---|---|---|---|
| Ω | 4 | 5 | 6 | 7 | 8 | 4 | 5 | 6 | 7 | 8 | 4 | 5 | 6 | 7 | 8 | 4 | 5 |
| correlated | 1.25 | 0.48 | 4.77 | 0.66 | undef. | 0.53 | 0.51 | 0.55 | 0.71 | 4.05 | 0.57 | 0.98 | undef. | 1.25 | 1.83 | 0.51 | 0.55 |
| random | 0.61 | 0.53 | 1.57 | 1.11 | 0.60 | 0.69 | 1.23 | 1.90 | 0.82 | 0.87 | 1.14 | 2.12 | 2.22 | 2.25 | 3.52 | 0.88 | 1.27 |

As we show above, the inclusion of GA-SISSO models obtained for lower complexity as input features into GA-SISSO jobs done for higher complexity (hiSISSO) improves prediction accuracy (Fig. 3 right). So, we applied this methodology also in the search for models of $CO_2$ adsorption energy. In all discussed cases we included GA-SISSO models with $D^{hiSISSO}$ = 4 or 5 and $Ω^{hiSISSO}$ = 2 as input feature into GA-SISSO models with $D$ = 4 or 5 and $Ω$ = 4 or 5. Applying genetic algorithm, the GA-SISSO model with the lowest 10-fold CV RMSE was obtained for $D$ = 5 and $Ω$ = 4 (with $D^{hiSISSO}$ = 5, $Ω^{hiSISSO}$ = 2). Its fitting RMSE is 0.33 eV, CV-RMSE and CV-MAE are 0.35 and 0.28 eV respectively (Figure 4 top). So, the expression for prediction of the $CO_2$ adsorption energy is:

$$E_{ads} = 0.63 + 1.01 \cdot 10^{-5} \alpha_{max} \cdot \Delta \cdot CBm^3 + 76.30 \cdot (\alpha_{max} \cdot U)^{-1} - 0.083 \cdot |\Delta\varphi \cdot U - \varphi_{1.4} \cdot E_1| \\ - 0.0013 \cdot \alpha_O \cdot \varphi_{2.6} \cdot U^3 - 1.12 \cdot ||\varphi_{1.4} - E_1| - |\varphi_{1.4}|| \qquad (2)$$

with $\alpha_{max}$ – polarizability of surface cations, maximal among all surface cations species, $\alpha_O$ – polarizability of surface O-atom above which $CO_2$ is adsorbed, $\Delta$ – band gap, CBm – conduction band minimum, $U$ – highest energy of 2p-band of O-atom, $\varphi_l$ – electrostatic potential above O-atom at the distance $l$ = 1.4 or 2.6 Å, $\Delta\varphi$ – difference of electrostatic potentials at 1.4 and 2.6 Å above O atom, and $E_1$ equal to:

$$E_1 = 2.38 - 0.37 \cdot (r_{max}^{LUMO} + d_1) + 0.03 \cdot d_3^3 - 15.44 \cdot E_{form}/r_{max}^{LUMO} + 0.60 \cdot |r_{max}^{LUMO} - d_1| + 7.62 \cdot d_3/PC \qquad (3)$$

here $r_{max}^{LUMO}$ is radius of LUMO of a gas-phase cationic atom, maximal among all cationic species, $d_l$ – distance from O-atom to the nearest $l$th cation, $E_{form}$ – surface formation energy. The 10-fold CV RMSE for $E_1$ model is 0.40 eV which is larger than for model (2).

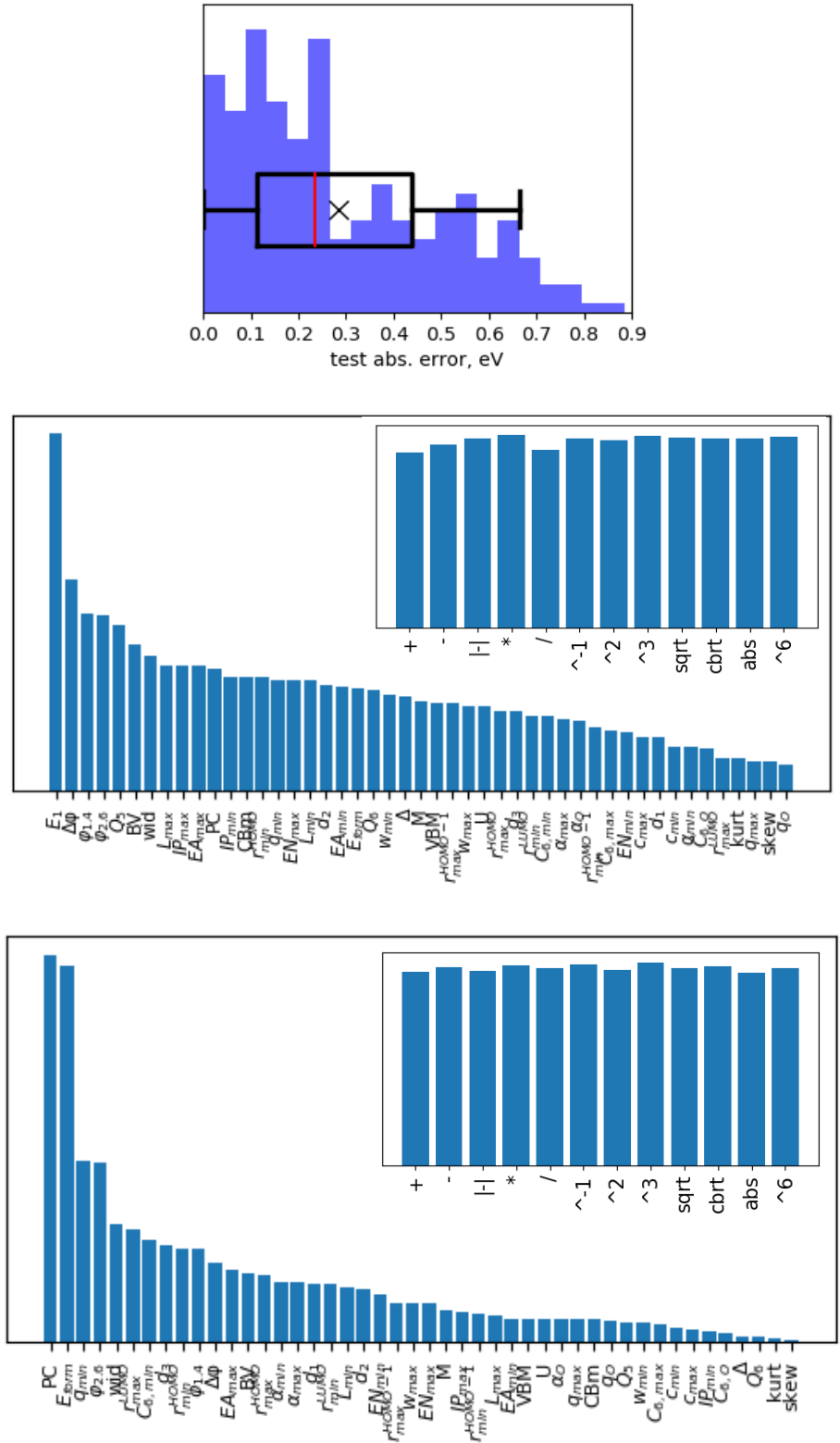

Figure 4. (top) Distribution of 10-fold CV absolute errors for SISSO model (2) for $CO_2$ adsorption energy; significance of primary features and operators (insets) according to GA performed within a hiSISSO approach: for GA-SISSO models obtained at the second step with $D = 5$ and $\Omega = 4$ (middle), and for the first step GA-SISSO models with $D^{hiSISSO} = 5$ and $\Omega^{hiSISSO} = 2$ (bottom).

Model (2) contains five terms. Among them, the one which explains the most of the variance of the adsorption energy is the last one: $\|\varphi_{1.4} - E_1| - |\varphi_{1.4}\|$. Its $R^2$ coefficient in fitting the adsorption energy is 0.74, much higher than for all other terms which are 0.21, 0.001, 0.05, and 0.03, respectively. This term besides $E_1$ contains also the electrostatic potential at 1.4 Å above surface O atom. 1.4 Å is also the maximal C-O bond distance between chemisorbed $CO_2$ molecule and the underlying surface oxygen atom. The electrostatic potential above surface O atoms is typically positive. This means that attraction of positively charged species such as carbon atom in a $CO_2$ molecule is favorable. Indeed, there is a certain relation between $\varphi_{1.4}$ and $CO_2$ adsorption energy (Figure S), although the corresponding $R^2$ correlation coefficient is only 0.36. The latter clearly indicates that attraction of positively charged carbon atom to surface oxygen is by far not the only factor influencing the $CO_2$ binding. As mentioned above, a very important role is played by the interaction of O atoms in $CO_2$ molecule with surface cationic atoms.

In the case of $E_1$ model, the main term is the ratio $d_3/PC$. The role of O 2p-band center is clear – the less negative its value is, the more charge is transferred to adsorbate molecule, the more negative is the bonding energy. Regarding the distance from surface oxygen to a third nearest cation ($d_3$), its larger values promote the strengthening of bonding. The reason is the ionicity. Larger interatomic distances in oxides imply larger radii of cations which in turn implies smaller values of absolute ionization potentials. Therefore, oxygen atoms in the neighborhood of such cations are expected to accumulate more charge density which can be further transported to a molecule adsorbed above. In other words, surfaces with larger oxygen-cation distances are more ionic, and this promotes the strengthening of bonds with $CO_2$.

For assessment of the importance of other features in description of $CO_2$ adsorption energies, we performed the analysis of all SISSO strings in the whole GA pool that where generated during the run. In this analysis, for each primary feature $k$ we calculated the so called significance (weighted frequency of appearance of a given feature or operator in the models along the GA):

$$s_k = \sum_{i}^{N_{all}} f_i \cdot 1\{k \in i\}$$

(4)

here 1{} is the indicator function which is 1 for *True* and 0 otherwise.

As expected, the most important feature is $E_1$ (Figure 4 middle). Among other features, the electrostatic potentials above oxygen atom at 1.4, 2.6 Å and especially their difference have highest significance. The difference $\Delta\varphi = \varphi_{1.4} - \varphi_{2.6}$ defines the direction of the electric field above the adsorption site, and in most of cases it is directed towards the surface. In cases where it is directed away from the surface, or towards the surface but has relatively small value, chemisorbed $CO_2$ structures have very weak binding energies, or even they are less stable than the physisorbed $CO_2$ structures. In general, there is certain relation between $CO_2$ adsorption energy and $\Delta\varphi$, although the $R^2$ coefficient in this case is relatively small 0.43 – even less than for $PC$ (0.58).

Since $E_1$ was obtained also on top of the same set of primary features as the model (2), and it encloses some of them, we have also done the same analysis for GA run for $D^{hiSISSO}$ = 5 and $\Omega^{hiSISSO}$ = 2 (Figure 4 bottom). Here two most significant features are O 2$p$-band center and surface formation energy ($E_{form}$), also $\varphi_{2.6}$ and minimal Hirshfeld charge of surface cations ($q_{min}$) have relatively high significances. The $PC$ and $E_{form}$ were found to be less significant in the case of the second step of hiSISSO for $D$ = 5 and $\Omega$ = 4 (Figure 4 middle), and they are not present in model (2). The reason is that on one hand they are already incorporated in $E_1$, and on the other hand $PC$ and $\Delta\varphi$ are correlated between each other ($R^2$ = 0.82). So, $PC$ is in some sense switched off in the second step of GA-hiSISSO. The surface formation energy should be related to adsorption energy in a way that less stable surfaces contain larger amount of chemically unsaturated atoms following stronger binding with adsorbate molecules. Despite that, we do not observe any correlation between $E_{form}$ and $CO_2$ adsorption energies (Figure S): the $R^2$-coefficient is 0.001. We have analyzed about thirty $\Omega^{hiSISSO}$ = 2 top strings in GA pool with lowest CV errors including the best one $E_1$, and we observed that $E_{form}$ appears in most of the cases during the second SISSO iteration while fitting the residual between the adsorption energy and the first term which contains $PC$. Thus, $E_{form}$ plays complementary role to $PC$ in the fitting $CO_2$ adsorption energies.

Regarding mathematical operators (insets in Figure 4), in both cases their significances $s_k$ according to (4) are always very close, with no evident outstanders in contrast to features. This suggests a weak dependence of final results on the operators sets. However, since studied operators set contains two different kinds of operators – unitary (powers, roots, absolute value) and binary ones ('+', '-', '*', '/', '|-|'), this requires probably another way of operators analysis with separation into two categories. This will be investigated later.

The SISSO model (2) appeared in 60 % of the cross-validation iterations. Although some terms inside this models are more persistent: the most important term $||\varphi_{1.4} - E_1| - |\varphi_{1.4}||$ was selected in all CV cases, the terms $(\alpha_0 \cdot \varphi_{2.6} \cdot U^3)$ and $|\Delta\varphi \cdot U - \varphi_{1.4} \cdot E_1|$ – in 90%, and the term $(\alpha_{max} \cdot U)^{-1}$ – in 70 %. With that we conclude that the found model is quite stable.

The prediction accuracy of the SISSO model about 0.3 eV is obviously larger than the accuracy of the used DFT approach for generation of the training set (~0.1 eV), and it is larger than one would need for example for predictions of catalytic materials in methane oxidation (0.2 eV) [25]. This is explained by the fact that applied approach provides a universal model which covers all possible mechanisms of $CO_2$ activation – by means of charge transport from underlying oxygen, and via one or two additional chemical bonds with surface cations. For establishing of more accurate models one can apply for example the subgroup discovery [23,31], or other alternative machine learning approaches which in turn might require the generation of a larger data set [32]. Also for some particular tasks active learning can be applied [25].

**Conclusions**

In this work we have developed a method for selection of primary features and mathematical operators in symbolic regression SISSO. The choice of an optimal set is important for avoiding the underfitting and overfitting of the obtained model. This features-selection method is genetic algorithm, and it deals with strings consisting of subsets of features and operators. We tested it for the set of perovskite materials with lattice vector as the target property [7]. We observed that for outperforming the model which is obtained while treating all features and operators simultaneously (in cases when it is not restricted due to computational limitations) in terms of prediction accuracy, important is the mutation of unselected during crossover features and operators as well as the inversion of the fitness-function. We found that this improvement is robust, i.e., it does not depend on the dimensionality and complexity of the SISSO model. Compared to SISSO models obtained from all features and operators we observed the improvement of prediction accuracy by 10 to 20 %.

Next we have applied our GA algorithm for the search of the SISSO model for prediction of $CO_2$ adsorption energies on semiconductor oxides [23]. The model obtained in this way has prediction accuracy about 0.3 eV that is much better than prediction accuracy of discussed in the literature models containing only O $2p$-band center. The analysis of obtained GA-SISSO model as well as of the GA performance showed that the main factors influencing the strength of $CO_2$ binding are electrostatic potentials above the surface oxygen atom, their difference, O $2p$-band center and surface formation energy. Although the latter has low correlation with adsorption energy, its role lies in fitting the residuals between $CO_2$ adsorption energies and O $2p$-band center. The analysis of mathematical operators showed a weak dependence of GA results on the operators sets, although this might need further investigations.

**References**

1. Y. Wang, N. Wagner, J.M. Rondinelli. MRS Commun. 2019, 9, 793–805.
2. D.A. Augusto, H.J.C. Barbosa. Proceedings - Brazilian Symposium on Neural Networks, IEEE Computer Society, 2000. 173.
3. L. Foppa, L.M. Ghiringhelli, F. Girgsdies, M. Hashagen, P. Kube, M. Hävecker, S.J. Carey, A. Tarasov, P. Kraus, F. Rosowski, R. Schlögl, A. Trunschke, M. Scheffler. MRS Bulletin. 2021, 46, 1016.
4. Xie, Stephan R., Gregory R. Stewart, James J. Hamlin, Peter J. Hirschfeld, and Richard G. Hennig. "Functional form of the superconducting critical temperature from machine learning." Physical Review B 100, no. 17 (2019): 174513.
5. Cao, Guohua, Runhai Ouyang, Luca M. Ghiringhelli, Matthias Scheffler, Huijun Liu, Christian Carbogno, and Zhenyu Zhang. "Artificial intelligence for high-throughput discovery of topological insulators: The example of alloyed tetradymites." Physical Review Materials 4, no. 3 (2020): 034204.
6. Ouyang, R.; Curtarolo, S.; Ahmetcik, E.; Scheffler, M.; Ghiringhelli, L. M. Phys. Rev. Mater. 2018, 2, 083802.
7. L. Foppa, T.A.R. Purcell, S.V. Levchenko, M. Scheffler, L.M. Ghringhelli. PRL 2022. **129**, 055301.
8. C. J. Bartel, S.L. Millican, A.M. Deml, J.R. Rumptz, W. Tumas, A.W. Weimer, S. Lany, V. Stevanović, C.B. Musgrave, A.M. Holder. Nat. Commun. 2018, 9, 4168.
9. M. Andersen, S. Levchenko, M. Scheffler, K. Reuter. ACS Catal. 2019, 9, 4, 2752.
10. Han, Z.-K.; Sarker, D.; Ouyang, R.; Mazheika, A.; Gao, Y.; Levchenko, S. V. Nat. Commun. 2021, 12, No. 1833.
11. R. Ouyang, E. Ahmetcik, C. Carbogno, M. Scheffler, L.M. Ghiringhelli. J. Phys.: Mater. 2019, 2, 024002.
12. T.A.R. Purcell, M. Scheffler, L.G. Ghiringhelli. J. Chem. Phys. 2023, 159, 114110
13. J. Fan, J. Lv. J.R. Statist. Soc. B. 2008, 70, 849-911.
14. C. Strobl, A.-L. Boulesteix, T. Kneib, T. Augustin, A. Zeileis. BMC Bioinformatics 2008, 9, 307.S
15. L.M. Ghiringhelli, J. Vybiral, S.V. Levchenko, C. Draxl, M. Scheffler. PRL 2015, 114, 105503.
16. C. Ding, H.C. Peng. *J. Bioinf. Comp. Biolog.* 2005, 03, 185.
17. B. Regler, M. Scheffler, L.M. Ghiringhelli. Data Min. Know. Disc. 2022, **36,** 1815.
18. Z. Guo, S. Hu, Z.-K. Han, R. Ouyang. J. Chem. Theory Comput. 2022, 18, 4945.
19. S. Bhattacharya, S.V. Levchenko, L.M. Ghiringhelli, M. Scheffler. PRL 2013, 111, 135501.
20. X. Zhao, X. Shao, Y. Fujimori, S. Bhattacharya, L.M. Ghiringhelli, H.-J. Freund, M. Sterrer, N. Nilius, S.V. Levchenko. J. Phys. Chem. Lett. 2015, **6**, 1204.
21. D.E. Goldberg. Genetic Algorithms in Search, Optimization, and Machine Learning. Addison-Wesley, 1989.
22. S. Bhattacharya, S.V. Levchenko, L.M. Ghiringhelli, M. Scheffler. NJP. 16 (2014) 123016.
23. A. Mazheika, Y.-G. Wang, R. Valero, L. Ghiringhelli, F. Vines, F. Illas, S. Levchenko, M. Scheffler. *Nat. Commun.*, 2022, 13, 419.
24. Wang, H., Schmack, R., Sokolov, S., Kondratenko, E., Mazheika, A., Kraehnert, R. *ACS Catal.* 2022, 12, 15, 9325.
25. A. Mazheika, M. Geske, M. Müller, S.A. Schunk, F. Rosowski, R. Kraehnert. ArXiv:2211.08014.
26. M.-M. Millet, G. Algara-Siller, S. Wrabetz, A. Mazheika, F. Girgsdies, D. Teschner, F. Seitz, A. Tarasov, S.V. Levchenko, R. Schlögl, E. Frei. JACS 2019, 141, 2451.
27. L. Giordano, K. Akkiraju, R. Jacobs, D. Vivona, D. Morgan, Y. Shao-Horn. Acc. Chem. Res. 2022, 55, 298.
28. C.F. Dickens, J.H. Montoya, A.R. Kulkarni, M. Bajdich, J.K. Nørskov. Surf. Sci. 2019, 681,

122.
29. A. Grimaud, K.J. May, C.E. Carlton, Y.-L. Lee, M. Risch, W.T. Hong, J. Zhou, Y. Shao-Horn. Nat. Commun. 2013, 4, 2439.
30. A. Maibam, S. Krishnamurty, M.A. Dar. *Mater. Adv.*, 2022, 3, 592.
31. C. Sutton, M. Boley, L.M. Ghiringhelli, M. Rupp, J. Vreeken, M. Scheffler. Nat. Commun. 2002, 11, 4428.
32. V. Fung, G. Hu, P. Ganesh, B.G. Sumpter. Nat. Commun. 2021, 12, 88.